\documentclass[14pt,a4paper]{article}
\usepackage{amsmath}
\usepackage[dvips]{graphicx}
\usepackage{setspace}

\usepackage[numbers,compress]{natbib}%

\usepackage[margin=2.5cm]{geometry}
\usepackage{epsfig}
\usepackage{subfigure}
\usepackage[section]{placeins}
\title{\bf SELECTED PROPERTIES  OF OPTICAL SPATIAL SOLITONS IN PHOTOREFRACTIVE MEDIA AND THEIR IMPORTANT APPLICATIONS}

\date{3 October, 2014}
\author{ S. Konar$^{1}$  and Vyacheslav A. Trofimov$^{2}$\\
$^{1}$Department of Applied Physics, Birla Institute of Technology,
Mesra, Ranchi-835215,\\ Jharkhand, India\\ \\$^{2}$Faculty of Computational  Mathematics and Cybernetics,\\ Lomonosov Moscow State University,  Vorobyovy Gory, Moscow 119992, Russia.\\\\}

\begin{document}
\maketitle

\setcounter{page}{1}

\begin{onehalfspacing}
\begin{abstract}

 { Some selected important properties of photorefractive spatial solitons and their applications have been reviewed in the present paper. Using band transport model, the governing principle of photorefractive nonlinearity has been addressed and nonlinear dynamical equations of spatial solitons owing to this nonlinearity have been discussed. Mechanisms of formation of screening and photovoltaic solitons of three different configurations i.e., bright, dark and grey varieties have been examined. Incoherently coupled vector solitons due to single and two-photon photorefractive phenomena have been highlighted. Modulation instability, which is precursor to soliton formation has been also discused briefly. Finally possible applications of photorefractive spatial solitons have been highlighted.}
\\\\\\\\\\\\\

Fax No:\ +91-6512276052,\ Tel No:\ +91-6512275522,\ Mobile No:\
+91-9431326757,\\     Email of corresponding author:\
swakonar@yahoo.com,\
                skonar@bitmesra.ac.in
\end{abstract}\ \ \ \ \ \ \ \ \ \ \ \ \ \ \ \ \ \ \ \ \ \
\end{onehalfspacing}
\setcounter{page}{1}
\ \ \ \ \ \ \ \ \ \ \ \ \ \ \ \ \ \ \ \ \ \ \ \ \ \ \

\newpage
\tableofcontents

\newpage
\section{Introduction}
\setcounter{equation}{0}

\ Nonlinear optics has lead the way of several fundamental discoveries, elegant and fascinating one of these is the optical solitons. These solitons are extensively studied since last three decades not only because of their mathematical elegance but also due to applications in photonics and optical communications~\citep{1,2,3,4,5,6,7,8,9,10,11,12,13,14,15,16,17,18,19,20,21,22,23,24}. Optical beams and pulses spread during propagation due to self-diffraction or dispersion. This broadening can be arrested in optical nonlinear media resulting in stable optical beams or pulses whose spreading is exactly balanced by optical nonlinearity and these self trapped beams or pulses are  known as optical solitons~\citep{2,3,5,7}. Optical spatial solitons are self-trapped optical beams whose spreading due to diffraction is exactly balanced by optical nonlinearity induced self lensing mechanisms.The optical beam induces a refractive index change in the medium thereby creating an optical waveguide which subsequently guides the beam. Thus, the beam first creates an waveguide and then self-trapped and guided in the waveguide created by itself. An optical temporal soliton on the otherhand is a nonspreading optical pulse whose broadening due to group velocity dispersion is exactly balanced by the nonlinearity induced self phase modulation. More explicitly, the group velocity dispersion produces a chirp in the propagating pulse which is balanced by the opposite chirp created by nonlinearity. Exact balance of these opposite chirps results in the formation of optical temporal solitons.
\subsection{Optical  Solitons  or Optical Solitary Waves? }
    Way back in 1995,  Zabusky and Kruskal~\citep{13}  introduced the term solitons to reflect the  particle like properties of  stable self-trapped waves in nonlinear media. The motivation behind this was the fact that the shape of these self-trapped wavepackets remain intact even after collision. Historically, this term was reserved for those wavepackets which obey integrable partial differential equations that can be solved by inverse scattering theory~\citep{10,11,13}. Solitons, which are solutions of these equations, remain invariant even after collision since they undergo elastic collision. Optical beam and pulse propagation in Kerr nonlinear media are governed by partial differential equations which are integrable via inverse scattering technique, hence, optical solitons can be created in such media. However, most physical optical systems, for example photorefractive media,  possess non-Kerr types of nonlinearity in which optical beams or pulses are described by dynamical equations that are though partial differential equations but  not integrable by inverse scattering technique. Self-trapped solutions of these non-integrable equations are known as 'solitary waves'. These solitary waves are not solitons since collision between these solitary waves is not always elastic~\citep{8,9,14}. For example, under specific conditions, two  photorefractive optical spatial solitary waves may coalesce to form a single solitary wave  and a single solitary wave might undergo fission to give birth to new solitons. However, these solitary waves in many cases play extremely important role in several applications.  In literature, despite their shortcoming, it is now quite common to  loosely refer all self-trapped optical solitary waves as solitons regardless of whether they obey integrable partial differential equation or not.
\subsection{Emergence of Photorefractive Optical Spatial Solitons }
    Guiding of optical beams in the self created waveguide was first pointed out by Askaryan in 1962~\citep{12}. Kelly~\citep{15} pointed out that in Kerr nonlinear media spatial solitons with two transverse dimensions(2D) are unstable and undergo catastrophic collapse. Soon it was realised that spatial solitons with only one transverse dimensions(1D) are stable in  Kerr nonlinear media~\citep{15}. Subsequently, it was also realized that catastrophic collapse of 2D solitons could be avoided in an optical media  possessing saturating nonlinearity~\citep{16}. Therefore, the idea of saturating nonlinearity was the key to  the discovery of a plethora of optical spatial solitons.\ Inspite of the prediction of  formation of stable 2D solitons in
       saturating nonlinear  media, no progress was made in creating these solitons due to lack of identification of appropriate optical materials that possess saturating nonlinearity. The breakthrough was made  by Segev et al~\citep{17,18} in 1990 with the identification of non-Kerr type of nonlinearity in photorefractive media and observation of optical spatial solitons due to such nonlinearities.  Unlike Kerr solitons, the  dynamics  of photorefractive solitons, under a variety of photorefractive nonlinear effects, is governed by modified   nonlinear Schr$\ddot{o}$dinger equations. Photorefractive nonlinearity is non-instantaneous, typically nonlocal and inherently saturable and spatial solitons in these media  can be created at microwatt power level using very simple set up in the laboratory, while their formation time ranges from microseconds to minutes. \\\\  Photorefractive optical spatial solitons  possess several unique properties, for example unlike Kerr solitons which undergo elastic collision, collisions of these solitons are inelastic, more diverse and interesting. The inelastic collision between two photorefractive spatial solitons may  lead to soliton fusion or fission, with particle-like annihilation or birth of new solitons.  Another unique feature is that one can create an induced  waveguide using a  weak soliton beam  which subsequently can be used to guide another powerful beam at different wavelength at which the material is less photosensitive ~\citep{24,25}.  It has been demonstrated that  photorefractive  spatial soliton-induced waveguides can be used for device applications such as directional couplers and high efficiency frequency  converters ~\citep{26,27}. These unique features of photorefractive nonlinearity have opened up immense possibilities for new and novel devices, such as all optical switching and routing, steering, interconnects, parallel computing, optical storage etc~\citep{28,29,30,31} .
      They are also promising for experimental verification of theoretical models, since, they can be created at very
      low power. In the present article, we confine our discussion on the properties of photorefractive optical spatial solitons.
       This review is only  partial description of the development of  photorefractive optical spatial solitons, interested readers are referred to several excellent recent reviews~\citep{14,32,33,34,35}.
\section{Photorefractive  Materials }
Exposure of  a photorefractive (PR) material  with optical field of nonuniform intensity leads to the excitation of charge carriers of inhomogeneous density. These charge carriers then migrate due to drift or diffusion or both and creates  space charge field which subsequently modifies the refractive index of the material via either linear electro-optic or quadratic electro-optic effect ~\citep{30,31,35}. Change in the refractive index of certain electro-optic materials due to optically induced redistribution of charge carriers is known as photorefractive(PR) effect. Photorefractive effect has
potential applications in holography~\citep{34}, optical phase
conjugation~\citep{30,31,36}, optical signal processing and optical
storage~\citep{31,34,35}. Generally, PR  materials are classified in three different categories. Most commonly used PR materials are inorganic ferroelectrics,  such as  $LiNbO_{3}$, $KNbO_{3}$, $BaTiO_{3}$ etc. Recently,  large number of experiments on spatial solitons have been performed~\citep{37} in  centrosymmetric paraelectric potassium lithium tantalate niobate $(KLTN)$. Some selected semiconductors such as InP, GaAs,  CdTe etc., also show PR property with  large carrier mobility that produces fast dielectric response time, which is important for fast image processing. They have potential use in  fast holographic processing of optical information. The third category PR materials are polymers which shows strong PR effect~\citep{34,35} at  high  applied voltage. PR pattern can be erased easily in polymers by decreasing applied voltage.
\section{Protorefractive Optical Nonlinearity}
 The band transport model of Kukhtarev-Vinetskii~\citep{38} has been  widely used to describe the theoretical foundation of PR nonlinearity. In view of this model, we assume a PR medium with completely full valance band and an empty conduction band which is illuminated by an optical field of nonuniforn intensity. It  has both donor and acceptor centers, uniformly distributed, whose energy states lies somewhere in the middle of the band gap. The donor electron states are at higher energy in comparison to that of  acceptor energy states. The nonuniform optical field excites unionized donors and creates charge carriers. These charge carriers move to the conduction band where they are free to move, to diffuse or to drift under the combined influence of self generated and
external electric field and are finally trapped by the acceptors. During this process, some of the electrons are captured by ionized
donors and thus are neutralized, and finally a steady state is reached with the creation of internal space charge field, that can be evaluated using  donor ionization rate equation, electron continuity equation, current density $(J)$ equation, Poisson's  equation and the charge density($\rho$) equation~\citep{14,17,18,19, 20,21,22,23,24,39,40} which are as follows:
\begin{equation}\label{e1}
\frac{\partial N_{D}^{+}}{\partial
t}=(s_{i}I+\beta_{T})(N_{D}-N_{D}^{+})-\gamma _{R}N N_{D}^{+}, \
\end{equation}
\begin{equation}\label{e2}
\frac{\partial N}{\partial t}-\frac{\partial N_{D}^{+}}{\partial t}
=\frac{1}{e}\overrightarrow{\nabla }.\overrightarrow{J},
\end{equation}
\begin{equation}\label{e3}
\overrightarrow{J}=e N\mu \overrightarrow{E} +k_{B}T\mu
\overrightarrow{\nabla }N +
k_{p}s_{i}\left(N_{D}-N_{D}^{+}\right)I\vec{c},
\end{equation}
\begin{equation}\label{e4}
\overrightarrow{\nabla}\cdot\epsilon \overrightarrow{E}=\rho ,
\end{equation}
\begin{equation}\label{e5}
\rho
\left(\overrightarrow{r}\right)=e\left(N_{D}^{+}-N_{A}-N\right),
\end{equation} where $N $, $N_{D}$, $N_{A}$  and $N_{D}^{+}$  are  densities of  electron, donor, acceptor, and  ionized donor, respectively;
 $s_{i}$ is the photoexcitation cross section, $I$ is the intensity of  light  in terms of Poynting flux,  $\beta_{T}$  and  $\gamma _{R}$  are rate of thermal generation and  electron trap recombination coefficient, respectively. $T$, $e$  and $\mu$ are respectively electron's temperature,  charge and  mobility; $k_{B}$ is the Boltzmann constant, $k_{p}$ is the photovoltaic constant and $\vec{c}$ is the unit vector in the
direction of c-axis of the PR crystal. The current density  arises due to the drift, diffusion and photovoltaic effect; $E$ is the sum of externally applied  field and the generated space charge field $E_{sc}$. Most experimental investigations on PR solitons
have been performed using one dimensional waves, therefore, it is appropriate to find material
response in one transverse dimension (say x only). In the steady state equations (1)-(4) reduces to:
\begin{equation}\label{e6}
s_{i}(I+I_{d})(N_{D}-N_{D}^{+})-\gamma_{R}NN_{D}^+=0,
\end{equation}
\begin{equation}\label{e7}
\frac{\partial E_{sc}}{\partial
x}=\frac{e}{\epsilon_{0}\epsilon_{r}}(N_{D}^{+}-N_{A}-N),
\end{equation}
\begin{equation}\label{e8}
J=e N\mu E_{sc} +k_{B}T\mu \frac{\partial N}{\partial x} +
k_{p}s_{i}\left(N_{D}-N_{D}^{+}\right)I,
\end{equation}
\begin{equation}\label{e9}
\frac{\partial J}{\partial x}=0,
\end{equation}  where   $I_{d}
(=\frac{\beta_{T}}{s_{i}})$ is the  dark irradiance which is also the homogeneous intensity that controls the
conductivity of the crystal. Usually  $I$ is such that   $N\ll
N_{D}$,  $N\ll N_{A}$ and $N_{A}\ll N_{D}^{+}$. The  space charge field $E_{sc}$  under these approximations turns out to be
\begin{equation}\label{e10}
E_{sc}=E_{0}\frac{I_{\infty} +I_{D}}{I+I_{D}}+E_{p}\frac{I_{\infty}
-I}{I+I_{D}}-\frac{k_{B}T}{e}\frac{1}{I+I_{D}}\frac{\partial
I}{\partial x},
\end{equation}  where  $E_{o}$  and $E_{p}=\frac{k_{p} \gamma_{R} N_{A}}{e\mu}$  are the external bias field to the
crystal and photovoltaic field, respectively. In  the above derivation  we have assumed that the power
density of the optical field  and space charge field attain asymptotically constant values
 i.e., $I(x \rightarrow \pm\infty,z)=I_{\infty}$ and  $E_{sc}(x \rightarrow
\pm\infty,z)=E_{o}$. The refractive index change $\triangle n$ in a  noncentrosymmetric PR crystal~\citep{14,17,18,19,20, 21, 22, 23, 24, 33, 34, 35}, such as $BaTiO_{3}$, $LiNbO_{3}$ etc., is due to the linear electro-optic effect (Pockel's) and is given
by
\begin{equation}\label{e11}
\Delta n =-\frac{1}{2}n^{3}r_{e}E_{sc},
\end{equation}
\begin{equation}\label{e12}
 =-\frac{1}{2}n^{3}r_{e}\left[E_{0}\frac{I_{\infty}
+I_{D}}{I+I_{D}}+E_{p}\frac{I_{\infty}
-I}{I+I_{D}}-\frac{k_{B}T}{e}\frac{1}{I+I_{D}}\frac{\partial
I}{\partial x}\right],
\end{equation} where $r_{e}$ and $n$  are  effective linear electro-optic coefficient and average refractive index, respectively. In  centrosymmetric PR materials the index change is due to the quadratic electro-optic response to a photoinduced internal field and can be expressed ~\citep{35,41} as
\begin{equation}\label{e13}
\Delta n =-\frac{1}{2}n_{b}^{3}g_{e}\epsilon_{0}^{2}(\epsilon_{r}-1)^{2}E_{sc}^{2} ,
\end{equation}
\begin{equation}\label{e14}
\triangle n =-\frac{1}{2}n_{b}^{3}g_{e}\epsilon_{0}^{2}(\epsilon_{r}-1)^{2}\left[E_{0}\frac{I_{\infty}
+I_{D}}{I+I_{D}}+E_{p}\frac{I_{\infty}
-I}{I+I_{D}}-\frac{k_{B}T}{e(I+I_{D})}\frac{\partial
I}{\partial x}\right]^{2},
\end{equation} where  $g_{e}$, $\epsilon_{r}$, $\epsilon _{0}$ and $n_{b}$  are the effective  quadratic electro-optic coefficient, relative dielectric constant, free space permittivity and  refractive index  of the crystal, respectively; it is assumed that the (dc) polarization is in the linear regime. Depending  on the sign of  $g_{e}$ the nonlinearity is either self-focusing or defocusing. The nonlinear property of the PR media is evident from  equations(12) and (14)  since the refractive index is intensity dependent. Different terms in the above expressions are responsible for the existence of different types of solitons. Screening and photovoltaic solitons respectively owe their existence to the first and second term in these expressions. When both first and second terms are dominant, one would expect screening photovoltaic solitons. The third term in each of these equations arises due to the diffusion process  and is not in general responsible for the formation of solitons, however, it is responsible for self deflection of solitons ~\citep{14,21,41}.
\section{ Spatial Optical Solitons Owing to Single Photon PR Phenomenon}
\subsection{Historical Development }
Way back in 1992, Segev et al~\citep{17,18} first detected PR solitons in quasi steady-state regime. Soon PR screening solitons were predicted and identified~\citep{19,20,21}. Two different varieties of steady state screening solitons were subsequently investigated and identified in different biased noncentrosymmetric media under varieties of experimental configurations~\citep{19,20,37,38,39,40,42,43,44,45}.
Subsequently, screening solitons were also predicted and observed in centrosymmetric materials, particularly in KLTN crystals~\citep{37,41}. Another family of optical solitons were also observed in unbiased PR crystals which exhibits photovoltaic effect. These solitons, popularly known as photovoltaic solitons, have been observed experimentally in 1D as well as 2D configurations~\citep{46,47,48}. Besides these, PR solitons which are combination of
the screening  and photovoltaic solitons were also predicted and successfully observed~\citep{42,49}. They owe their existence to both photovoltaic effect and spatially nonuniform screening of the applied field, and, are also known as screening photovoltaic (SP) solitons.  By controlling the magnitude of bias field they can be converted to screening solitons or photovoltaic solitons.\\\\  PR crystals like $LiNbO_{3}$ possess high second-order susceptibility, hence, can be used for parametric processes such as second-harmonic generation.  Recently, it has become very popular,  primarily due to its ultra slow PR relaxation time owing to which any written solitonic channels can act as optical waveguides for long time even after turning off the soliton beam. These soliton induced waveguides have been  used both for switching devices ~\citep{50} and  enhancing the second harmonic conversion efficiency ~\citep{51}. Though  non-centrosymmetric PR crystals possess large nonlinearity, their typical response time is in the range of  milliseconds, hence, these materials are not suitable for very fast reconfigurable waveguide channels and switches. On the otherhand using paraelectric centrosymmetric PR crystals, particularly KLTN, DelRe et al.~\citep{52} demonstrated  dynamic switching  using electroholography~\citep{52,53,54}. Spatial solitons are  created   in paraelectrics  that  in turn is employed to guide the signal beam of  non-photorefractive wavelength. The soliton induced waveguide remains stable even at high power levels, and can be modified  via the electro-optic effect  by varying external biasing field, ensuring  fast change of propagation properties of the signal beam. Fast electro-optic response of $ KLTN$ makes them useful in important applications like optical communications and signal processing and this has paved the way to fast nanosecond applications of PR  spatial solitons ~\citep{52,53,54}.
\subsection{Modified Nonlinear Schr$\ddot{o}$dinger Equation($\textbf{mNLSE}$)}
Most of the solitary wave experiments in PR media employ optical beams with one transverse dimension, therefore, we assume the optical
beam is such that no dynamics is involved in y-direction and it is permitted to
diffract only along $x$ direction. The external bias field $E_{0}$  and optic axis of the crystal are  directed along the $x$ axis.The extraordinary refractive
index $\hat{n}_{e}$  in such cases~\citep{14,20,42,55} is given by $({\hat{n}_{e})^{2}}=n_{e}^{2}-n_{e}^{4}r_{e}E_{sc},$  where $n_{e} $ is the unperturbed extraordinary index of refraction. The electric field of the soliton forming beam is assumed to be
$\overrightarrow{E}= \overrightarrow{x}\Phi(x,z)exp[i(kz-\omega
t)],$  where $k=k_{0}n_{e}$, $k_{0}=\frac{2\pi}{\lambda _{0}}$, $\lambda_{0}$  is the
free space wavelength of the optical field. Employing slowly varying envelope approximation for $\Phi $ in the Maxwell's equations
 we obtain
\begin{equation}\label{e15}
i\frac{\partial \Phi}{\partial
z}+\frac{1}{2k}\frac{\partial^{2}\Phi}{\partial
x^{2}}-\frac{1}{2}k_{0}n_{e}^{3}r_{e}E_{sc}\Phi=0.
\end{equation} By virtue of the  space charge field  evaluated  in section (3), we obtain following~\citep{14,21,69,70,71,72,73} modified nonlinear
Schr\"{o}dinger equation (mNLSE):

\begin{equation}\label{e16}
i\frac{\partial A}{\partial\xi }+\frac{1}{2}\frac{\partial^{2}
A}{\partial s^{2}}-\frac{\beta (1+\rho)A}{1+|A|^{2}}-
\frac{\alpha (\rho-|A|^{2})A}{1+|A|^{2}}+\frac{\delta A}{1+|A|^{2}}\frac{\partial
|A|^{2}}{\partial s}=0,
\end{equation} where $\xi=\frac{z}{k_{0}n_{e}x_{0}^{2}}$, $s=\frac{x}{x_{0}}$, $\alpha=(k_{0}x_{0})^{2}(n_{e}^{4}r_{e}/2)E_{p}$, $\beta=(k_{0}x_{0})^{2}(n_{e}^{4}r_{e}/2)E_{0}$,
$\rho=\frac{I_{\infty}}{I_{d}}$, $\delta=(k_{0}^{2}x_{0}r_{e}n_{e}^{4}k_{B}T)/(2e)$, and
$A=\sqrt{\frac{n_{e}}{2\eta_{0}I_{d}}}\Phi$. Above mNLSE is the governing equation for varieties of bright, dark and gray solitons.
Two parameters  $\alpha $  and $\beta$  paly very important role in the formation of these solitons, while  $\delta$, which is associated with the diffusion term, is not directly responsible for soliton formation. The diffusion processes is primarily responsible for bending of trajectories of
propagating solitons, hence, large value of $\delta$ influences the trajectory of bending.
\section{Screening Optical Spatial Solitons}
Screening solitons  could be created in biased nonphotovoltaic PR crystals (i.e., $\alpha$=0 ), hence, if we neglect diffusion then the equation for screening solitons reduces to
\begin{equation}\label{e17}
i\frac{\partial A}{\partial\xi }+\frac{1}{2}\frac{\partial^{2}
A}{\partial s^{2}}-\beta( 1+\rho)\frac{A}{1+|A|^{2}}=0.
\end{equation}  \textbf{Bright Screening Solitons:}  The soliton forming beam for bright solitons  vanishes at infinity, therefore $I_{\infty}=\rho=0$. We express the stationary bright soliton  as $A(s,\xi)=p^{\frac{1}{2}}y(s)\exp(i\nu \xi) $,  where $\nu$ is the
nonlinear shift in propagation constant and  y(s) is a normalized
real function. In addition, $0\leq y(s)\leq1$,  $y(0)=1$, $\dot{y}(0)=0$ and $y(s\rightarrow{\pm \infty})=0$. The parameter \emph{p} represents the ratio of the peak intensity $(I_{max})$  to the dark
irradiance $I_{d}$, where $I_{max}=I(s=0)$.  Substitution  for $A(s,\xi)$ into equation (17) yields
\begin{equation}\label{e18}
\frac{d^{2}y}{ds^{2}}-2\nu y-2\beta\frac{y}{1+py^{2}}=0.
\end{equation} Use of boundary condition and integration of above equation leads to
\begin{equation}\label{e19}
s=\pm\frac{1}{(2\beta)^{1/2}}\int_{y}^{1}\frac{p^{1/2}}{[\ln(1+p\widehat{y}^{2})-\widehat{y}^{2}\ln(1+p)]^{1/2}}
d\widehat{y}.
\end{equation} The  bright profile $y(s)$ can be obtained using numerical procedure and can be easily shown that these solitons exist when $\beta>0 $
i.e., $E_{0}$ is positive.\\\\ \textbf{Dark Screening Solitons:} In appropriate media, equation (17) also admits dark solitary wave solutions~\citep{14,21} which  are embedded in a constant intensity background, therefore,  $I_{\infty}$ is finite, hence, $\rho$ is
also finite.  In addition, they exhibit anisotropic field profiles with respect to $s$. We take following ansatz for stationary solutions: $A(s,\xi)=\rho^{1/2} y(s)\ \exp(i\nu\xi)$, where $\nu$ is the nonlinear shift in propagation constant and  $y(s)$ is
a  normalized real odd function of $s$ satisfying   $y(0)=0, y(s\rightarrow\pm
\infty)=\pm1, \frac{dy}{ds}=\frac{d^{2}y}{ds^{2}}=0$  as
$s\rightarrow \pm\infty$.\ Substituting  $A$ in equation (17) we obtain,
\begin{equation}\label{e20}
\frac{d^{2}y}{ds^{2}}-2\nu y-2\beta ( 1+\rho)\frac{y}{1+\rho
y^{2}}=0.
\end{equation} By virtue of integration  and use of boundary condition, we immediately obtain
\begin{equation}\label{e21}
s=\pm\frac{1}{(-2\beta)^{1/2}}\int_{y}^{0}\frac{d\widehat{y}}{[(\widehat{y}^{2}-1)-\frac{(1+\rho)}{\rho}\ln\frac{1+\rho
\widehat{y}^{2}}{1+\rho}]^{1/2}}.
\end{equation} Obviously, these solitons  exist only when $\beta<0 $
i.e., $E_{0}$ is negative. Unlike their bright counterpart, these dark screening photovoltaic solitons do not
possess bistable property. \\\\ \textbf{Gray Screening Solitons:} Besides bright and dark solitons, equation (17) also admits another
interesting class of solitary waves, which are known as gray solitons~\citep{14,21}. In this case too, wave power density attains a
constant value at infinity i.e., $I_{\infty}$  is finite, and hence,
$\rho$ is finite. To obtain stationary solutions, we assume
\begin{equation}\label{e22}
A(s,\xi)=\rho^{1/2}y(s)\exp[i(\nu\xi+\int^{s}\frac{Jd\widehat{s}}{y^{2}(\widehat{s})})],
\end{equation} where $\nu$ and $J$ are  nonlinear shift in propagation constant and a
real constant, respectively; $y(s)$ is a normalized real even function of $s$ with  properties  $y^{2}(0)=m\ (0<m<1)$,  $\dot{y}(0)=0$, $y(s\rightarrow
\pm\infty)=1$ and all derivatives of $y(s)$  are zero at infinity. The parameter  \emph{m} describes grayness, i.e., the intensity  $I(0)$ at
the beam center  is $I(0)=mI_{\infty}$.  Substitution of the above
ansatz for $A$ in equation (17) yields
\begin{equation}\label{e23}
\frac{d^{2}y}{ds^{2}}-2\nu y-2\beta ( 1+\rho)\frac{y}{1+\rho
y^{2}}-\frac{J^{2}}{y^{3}}=0.
\end{equation} The values of $J$ and $\nu$ can be obtained using  boundary conditions mentioned earlier. Inserting these values and after integrating once, we get
\begin{equation}\label{e24}
\left(\frac{dy}{ds}\right)^{2}=2\nu(y^{2}-1)+\frac{2\beta}{\rho}(1+\rho)\ln
\left(\frac{1+\rho y^{2}}{1 + \rho
}\right)+2(\nu+\beta)\left(\frac{1-y^{2}}{y^{2}}\right).
\end{equation}   Profiles of dark solitons can be obtained easily  by numerical
integration of above equation.  Unlike bright or dark
solitons, the phase of gray solitons is not constant across $s$,
instead varies across $s$ and they can exist only when $\beta<1$  and   $ m<1$.

\section{ Optical Spatial Vector Solitons }
In previous sections, our discussions were confined to optical spatial solitons which are solutions of a single dynamical equation. These solutions arise due  a single optical beam with specific polarization. However, there are instances where two or more optical beams may mutually get trapped and propagate without distortion. These beams could be of same or different frequencies or polarization. they are mutually trapped and depend on each other in such a way that undistorted propagation of one is sustained by other and vice versa. In order to describe self and mutually trapped propagation of more than one soliton forming optical beams, we need to solve a set of more than one coupled solitary waves. Solutions of this set of coupled NLS equations are called vector
solitons if they preserve their shape.

\subsection{Incoherently Coupled Spatial  Vector Solitons}
Steady state incoherently coupled solitons are most extensively studied vector solitons in PR media ~\citep{14,58,59,60,61,62,63,64,65,66,67,68,69,70}. These solitons exist only when the two soliton forming beams possess same polarization and frequency and are mutually incoherent. Four different varieties of solitons  i.e.,
 bright-bright, dark-dark, bright-dark and gray-gray~\citep{63,69,70}
have been observed. Since two beams are mutually incoherent, no phase
matching is required and they experience equal effective
electro-optic coefficients. The idea of two incoherently coupled
solitons has been generalized and extended to soliton families where
number of constituent solitons are more than two.

\subsection{Coupled  mNLSE Owing to  Single Photon Phenomenon}
  We consider a pair of mutually incoherent optical beams of same frequency and polarization which are
propagating in a lossless PR crystal  along \emph{z}-direction. The optical c-axis of the crystal and polarization of both beams are
oriented along the \emph{x}-direction. These  beams are
allowed to diffract only along the \emph{x}-direction and \emph{y}-dynamics has
been implicitly omitted in the analysis.  These beams are expressed as $\overrightarrow{E_{j}}
=\overrightarrow{x}\Phi_{j}(x,z)\exp(ikz)$, $j=1, 2 $ and  $\Phi_{j}$ is the  slowly varying envelope of
optical field which satisfies following  equation:
\begin{equation}\label{e25}
i\frac{\partial \Phi_{j}}{\partial
z}+\frac{1}{2k}\frac{\partial^{2}\Phi_{j}}{\partial
x^{2}}-\frac{k_{0}n_{e}^{3}r_{33}E_{sc}}{2}\Phi_{j}=0,
\end{equation} Neglecting  diffusion effect, the  space charge field  can be obtained from  equation (10) as
\begin{equation}\label{e26}
E_{sc}=E_{0}\frac{I_{\infty}+I_{d}}{I+I_{d}}+E_{P}\frac{I_{\infty}-I}{I+I_{d}},
\end{equation}  where  $I(x,z)=n_{e}/(2\eta_{0} )(|\Phi_{1}|^{2}+|\Phi_{2}|^{2})$ is total power density of two  beams, whose value at a distance far away from the center of the crystal is  $I_{\infty}=I(x\rightarrow \pm\infty)$.  Substituting the expression of  $E_{sc}$ in
equation (25), we derive following equation:
\begin{eqnarray}\label{e27}
i\frac{\partial A_{j}}{\partial\xi }+\frac{1}{2}\frac{\partial^{2}
A_{j}}{\partial s^{2}}-\beta(1+\rho)
\frac{A_{j}}{(1+|A_{j}|^{2}+|A_{3-j}|^{2})}\nonumber \\
-\alpha\frac{(\rho-|A_{j}|^{2}-|A_{3-j}|^{2})A_{j}}{(1+|A_{j}|^{2}+|A_{3-j}|^{2})}
=0,
\end{eqnarray} where $A_{j}=\sqrt{\frac{n_{e}}{2\eta_{0}I_{d}}}\Phi_{j}$;\ $\alpha$,  $\beta$, $\xi$, $s$ and  $\rho$ are defined earlier. Above set of two coupled Schr\"{o}dinger equations can be examined for bright-bright, bright-dark, dark-dark,
gray-gray screening, photovoltaic as well as screening photovoltaic
solitons~\citep{59,60,61,62,63,64,65,66,67,68,69}. In the theoretical front, numerical method to solve above set of coupled equations was developed by Christodoulides et. al.~\citep{58}. Though this method has been used extensively, it fails to identify the existance of large family of solitons. Konar et al~\citep{70} has developmed a method which captures those solitons missed out by Christodoulides et. al.~\citep{58}. In next few lines we describe the method   to identify bright-dark solitons only. For elaboration readers are referred to ~\citep{14,59,60,61,62,63,64,65,66,67,68,69,70}.\\\\  \textbf{Bright-dark soliton:} We express  $A_{1}=p^{1/2} f(s)\exp(i\mu\xi)$   and
  $A_{2}=\rho^{1/2} g(s)\exp(i\nu\xi)$, where $f(s)$  and $g(s)$  respectively represents envelope of bright and dark beams.
   Two positive quantities  $p$  and $\rho$ represent the ratios of their maximum power density with respect to the dark
   irradiance $I_{d}$. Therefore, bright-dark soliton pair obeys following coupled ordinary differential equations:
\begin{eqnarray}\label{e28}
\frac{d^{2}f}{ds^{2}}-2\left[\mu +
\frac{\beta(1+\rho)}{1+pf^{2}+\rho g^{2}}\right]f=0,
\end{eqnarray} and
\begin{eqnarray}\label{e29}
\frac{d^{2}g}{ds^{2}}-2\left[\nu +
\frac{\beta(1+\rho)}{1+pf^{2}+\rho g^{2}}\right]g=0.
\end{eqnarray}  A particular solution of above equations is obtained assuming   $f^{2}+g^{2}=1$. Use of  boundary
conditions gives  $\mu=-\frac{\beta}{\Lambda} \ln(1+\Lambda)$  and
$\nu=-\beta$, where $ \Lambda= (p-\rho)/(1+\rho)$. When peak
intensities of  two solitons are approximately equal ($\Lambda<<1$), the  soliton solution~\citep{58,71}  leads to  $A_{1}=p^{1/2}$\
sech$[(\beta\Lambda)^{1/2}s]\exp[-i\beta(1-\Lambda/2)\xi],$ and
$A_{2}=\rho^{1/2}\tanh[(\beta\Lambda)^{1/2}s]\exp[-i\beta\xi],$ which exist only when $(\beta\Lambda >0)$.

\section{Two-Photon Photorefractive Phenomenon}
In previous sections, we have discussed properties of optical spatial solitons which owe their existence due to single
 photon PR phenomenon. Recently, Ramadan et al.~\citep{72} created bright spatial solitons in a biased $BSO$ crystal using  two-step excitation process.  Electrons were first excited to the conduction band by a background beam, and then they were excited towards higher levels in the conduction band by a second  optical beam of larger wavelength. Using this two-step process in a biased  $BSO$  crystal, Ramadan et al.~\citep{72} demonstrated the self-confinement of a red beam  at  633 nm  supported by another optical beam at 514.5 nm. Recently,  Castro-Camus and Magana~\citep{73} also presented an identical model of two-photon PR phenomenon which includes a valance band (VB), a conduction band (CB) and an intermediate allowed level (IL). A gating beam of  photon  energy  $\hbar\omega_{1}$ is used to maintain a quantity of excited electrons from the valance band (VB) to an intermediate allowed level (IL) which are subsequently  excited to the conduction band by another signal beam with photon energy $\hbar\omega_{2}$. The   signal beam  can induce a  spatial dependent charge distribution leading  to a nonlinear change of refractive index in the medium. Based on Castro-Camus and Magana's model, several authors have investigated two-photon screening  and   photovoltaic solitons ~\citep{10,14,68,74,75} which owe their existence due to two-photon PR phenomenon.

\subsection{Optical Nonlinearity and Evolution Equation of Solitons  }

In order to estimate optical nonlinearity owing to two-photon PR phenomenon, we need to evaluate the space charge field which can be obtained from the set of rate, current and Poisson's equations proposed by Castro-Camus et al.~\citep{73}. Instead of doing that here we refer interested readers to references~\citep{14,75} and straightaway use the expression of space charge field $E_{sc}$ due to
two-photon PR phenomenon~\citep{75}. We assume the soliton forming optical beam of intensity $I_{2}$  is propagating  along the $z$ direction of the crystal which is permitted to diffract only along the $x$ direction. The optical beam is polarized along the $x$ axis which is also the direction  crystal c-axis and the external bias field. The soliton forming beam  is taken as $\overrightarrow{E}=\overrightarrow{x}\Phi(x,z)\exp[i(kz-\omega t)]$, where the symbols have been defined earlier. The crystal is biased with external voltage $V$ and connected with external resistance $R$ and it is   illuminated with a gating beam of
constant intensity $I_{1}$. We assume that both the power density  and space charge field are uniform at large distance from the
center of the soliton forming beam, thus,  $ I_{2} (x\rightarrow \pm\infty, z)=I_{2\infty}$ =constant and  $E_{sc} (x\rightarrow
\pm\infty, z)=E_{0}$. Neglecting the effect of diffusion, the space charge field turns out to be
\begin{eqnarray}\label{e30}
E_{sc}&=&gE_{a}\frac{(I_{2\infty
}+I_{2d})(I_{2}+I_{2d}+\frac{\gamma_{1}N_{A}}{s_{2}})}{(I_{2
}+I_{2d})(I_{2\infty}+I_{2d}+\frac{\gamma_{1}N_{A}}{s_{2}})} \nonumber
\end{eqnarray}
\begin{eqnarray}\label{e30}
+E_{p}\frac{s_{2}(gI_{2\infty
}-I_{2})(I_{2}+I_{2d}+\frac{\gamma_{1}N_{A}}{s_{2}})}{(I_{2
}+I_{2d})(s_{1}I_{1}+\beta_{1})},
\end{eqnarray}  where $N_{A}$ and $N$ are acceptor  density and electron density in conduction band, respectively; $\gamma_{R}$, $\gamma_{1}$ and  $\gamma_{2}$ are  the recombination factor of the conduction to valence band transition, intermediate allowed level to valence band transition  and   conduction band to intermediate level transition, respectively; $\beta_{1}$ and $\beta_{2}$ are respectively the thermo-ionization probability constant for transitions from valence band to intermediate level and intermediate level to conduction
band; $s_{1}$ and $s_{2}$ are photo-excitation crosses. $E_{p}=\kappa_{p} N_{A}\gamma_{R}/e\mu $ is the
photovoltaic field, $I_{2d}=\beta_{2}/s_{2}$  is the  dark
irradiance,  $g= 1/(1+q)$, $q= \frac{e\mu N_{\infty}SR}{d}$,
$N_{\infty}=N( x\rightarrow \pm\infty )$. In general, $g$  is bounded between $0\leq g\leq 1$.
Under short circuit condition  $R=0$ and $g=1$, implying  the
external electric field is totally applied to the crystal. For open
circuit condition $R\rightarrow \infty,$  thus,   $g=0$ i.e., no
bias field is applied to the crystal. Employing equation (30) and
following the procedure employed earlier, the nonlinear
Schr\"{o}dinger equation for the normalized envelope can
be obtained as~\citep{75,77}
\begin{eqnarray}\label{e31}
i\frac{\partial A}{\partial\xi }+\frac{1}{2}\frac{\partial^{2}
A}{\partial s^{2}}-\beta g\frac{(1+\rho)(1+\sigma+|A|^{2})A}{(1+|A|^{2})(1+\sigma+\rho)}\nonumber \\
-\alpha\eta\frac{(g\rho-|A|^{2})(1+\sigma+|A|^{2})A}{1+|A|^{2}}=0,
\end{eqnarray} where   $\rho= I_{2\infty}/I_{2d}$,    $A=
\sqrt{\frac{n_{e}}{2\eta_{0}I_{2d}}}\Phi$,   $\beta=(k_{0} x_{0} )^{2} (n_{e}^{4}
r_{33}/2) E_{a}$, $\eta=\beta_{2}/(s_{1} I_{1}+\beta_{1}),
\sigma=\frac{\gamma_{1}N_{A}}{s_{2}I_{2d}}=\frac{\gamma_{1}N_{A}}{\beta_{2}}$.
Equation (31) can be employed to investigate screening, photovoltaic
and screening photovoltaic solitons under appropriate experimental
configuration. \\\\  \textbf{Bright Screening Solitons:} These solitons, for which $\alpha =0$  and $\rho =0 $, have been studied by several authors~\citep{14,73,74,75,76,77}.  In the low amplitude limit  i.e., when  $|A|^{2}<<1$, equation (31)  reduces to
\begin{equation}\label{e32}
i\frac{\partial A}{\partial\xi }+\frac{1}{2}\frac{\partial^{2}
A}{\partial
s^{2}}-\frac{\beta}{1+\sigma}\left(1+\sigma-\sigma|A|^{2}\right)A
=0.
\end{equation}  The one-soliton solution  of above equation is given by
\begin{equation}\label{e36}
A(s,\xi)=p^{1/2}sech\left[\left(\frac{\beta p
\sigma}{1+\sigma}\right)^{1/2}s\right]\\exp\left[i\frac{\beta(p\sigma-2\sigma-2)}{2(1+\sigma)}\xi\right].
\end{equation}  \\\ \textbf{Dark screening solitons:} For this case $\rho \neq0 $, thus when $|A|^{2}<<1$, \ equation (31) reduces to
\begin{equation}\label{e37}
i\frac{\partial A}{\partial \xi}+\frac{1}{2}\frac{\partial^2
A}{\partial
s^2}-\frac{\beta(1+\rho)}{(1+\sigma+\rho)}(1+\sigma-\sigma |A|^2)A
=0.
\end{equation} The dark soliton solution of above equation turns out to be \begin{equation}\label{e38}
A(s,\xi)=\rho^{1/2}\tanh\left[-\left(\frac{\beta\rho\sigma}{1+\sigma+\rho}\right)^{1/2}s\right]exp\left[\frac{i\beta(1+\rho)(\rho\sigma-\sigma-1)\xi}{1+\sigma+\rho}\right].
\end{equation}  These solitons could be observed in  $SBN$  since, they have an intermediate level required for two step
excitation. In addition to bright and dark solitons, equation (31)
also predicts steady state gray solitons which were investigated by Zhang
et. al.~\citep{76}. Characteristics of these solitons  are similar
to  one-photon PR gray spatial solitons. For example, they require bias field in opposite to the
optical c-axis and their FWHM is inversely proportional to the square root of the absolute value of the bias
field.  Proceeding in a similar way we can study bright and dark photovoltaic solitons.
\section{ Modulation Instability(MI)}
Modulation instability (MI) is an inherent characteristic of wave
propagation in nonlinear media. It refers to unstable
propagation of a continuous or quasi-continuous wave(CW) in such a
way that the wave disintegrates into large number of localized
coherent structures after propagating some distance through
nonlinear physical media. MI occurs as a result of interplay between nonlinearity and dispersion in temporal domain,
and between nonlinearity and diffraction in the spatial domain~\citep{2,5,81,82,83,84,85,86,87,88,89,90}. A continuous wave (CW) or quasi-continuous wave radiation propagating in a nonlinear medium may suffer instability with
respect to weak periodic modulation of the steady state and results
in the breakup of CW into a train of ultra short pulses.  In spatial
domain, for a narrow beam, self phase modulation exactly balances
the diffraction and a robust spatial soliton is obtained, while a
broad optical beam disintegrates into many filaments during
propagation in the same self-focusing nonlinear medium.\\\\\
MI has been made extensively studied in a wide range of
physical systems like fluids~\citep{79}, plasmas ~\citep{81}, Bose Einstein
condensates~\citep{82}, discrete nonlinear systems~\citep{83},
negative index materials~\citep{84}, soft matter~\citep{85}, PR media~\citep{86,87}, optical fibers
~\citep{88} etc.  MI typically occurs in the same parameter region where  solitons are observed. In fact, the filaments that emerge from the MI process
are actually trains of almost ideal solitons. Therefore, the
phenomenon of MI can be considered as a precursor of soliton
formation and has been found in both coherent beams and
incoherent beams. MI  has been extensively investigated in PR media ~\citep{80,90,91}. Single as well as two-photon PR media have been considered to analyse instability characteristics. Moreover, not only noncentrosymmetric but centrosymmetric
PR media have been examined as well ~\citep{90}. Unlike
noncentrosymmetric media, in a centrosymmetric media the
characteristics of this instability is independent of the external
applied field. In the next section, we examine the MI of a broad
optical beam in a biased two-photon non-centrosymmetric photovoltaic PR medium.

\subsection{ MI Gain under Linear Stability Framework}
In order to find out MI gain, we consider an  optical beam with large transverse spatial dimension.  Since we are confining our present
interest on the stability of a broad bright beam of finite transverse
extension, therefore   $I_{2\infty}=0$   and hence, $\rho=0$.
Therefore, the evolution equation of the broad optical field reduces
to
\begin{eqnarray}\label{e36}
i\frac{\partial A}{\partial\xi }+\frac{1}{2}\frac{\partial^{2}
A}{\partial s^{2}}-\beta
g\frac{(1+\sigma+|A|^{2})}{(1+|A|^{2})(1+\sigma)}A\nonumber \\
+\alpha\eta\frac{|A|^{2}(1+\sigma+|A|^{2})}{(1+|A|^{2})}A=0.
\end{eqnarray}  Equation (36) admits a steady state CW solution $A(\xi,s)=\sqrt{P}\exp[i\Psi(\xi)],$ where $P$  is the initial input power at $\xi=0$  and $\Psi(\xi)$ is the nonlinear phase shift which increases with propagation distance $\xi$ according to
\begin{eqnarray}\label{e37}
\Psi(\xi)=-g\beta\frac{(1+\sigma+P)}{(1+P)(1+\sigma)}\xi+\alpha\eta\frac{P(1+\sigma+P)}{1+P}.
\end{eqnarray}  The initial stage of MI can be investigated
in the linear stability framework, under which the stability of the steady state
solution is examined by introducing a perturbation in the amplitude
of the beam envelope so that the perturbed field now becomes:
\begin{eqnarray}\label{e38}
A(\xi,s)=\left[\sqrt{P}+a(\xi,s)\right]\exp[i\Psi(\xi)],
\end{eqnarray}  where $a(\xi,s)$ is an arbitrarily small complex perturbation field
such that  $a(\xi,s)<<\sqrt{P}$. Substituting  the perturbed field in  equation (36) and retaining terms linear only in the perturbed
quantity, the evolution equation for the perturbation field is obtained as
\begin{equation}\label{e39}
i\frac{\partial a}{\partial \xi}+\frac{1}{2}\frac{\partial
^{2}a}{\partial s^{2}}+\alpha\eta
P(a+a^{*})+\frac{\alpha\eta\sigma P(a+a^{*})}{(P+1)^{2}}+\frac{g\beta\sigma P(a+a^{*})}{(1+\sigma)(1+P)^{2}}=0
\end{equation} where $*$ denotes complex conjugate. The spatial perturbation
$a(\xi,s)$ is assumed to be composed of two sideband plane waves
\begin{equation}\label{e40}
a(\xi,s)=u\cos(K\xi-\Omega s)+iv \sin (K\xi-\Omega s),
\end{equation} where  $u $   and    $v$  are the real amplitudes of the perturbing
field, $K$ and $\Omega$  being the wave number and spatial frequency
of the perturbations, respectively. Substitution for the
perturbation field into its evolution equation yields:
\begin{equation}\label{e41}
\begin{pmatrix}
\Pi^{-} & -K\\
K & \Pi^{+}\\

\end{pmatrix}
\begin{pmatrix}
u\\
v\\

\end{pmatrix}
=
0\\,
\end{equation}   where $\Pi^{-}=-\Pi^{+}+\frac{2g\beta\sigma
P}{(1+\sigma)(1+P)^{2}}+\frac{2\alpha\eta\sigma
P}{(1+P)^{2}}+2\alpha\eta P$   and   $\Pi^{+}=\Omega^{2}/2$.
Equation(40) possesses a nontrivial solution only when the
following dispersion relation holds good:
\begin{equation}\label{e42}
K^{2}=-\frac{\Omega^{2}}{2}\left(\frac{2g\beta\sigma
P}{(1+\sigma)(1+P)^{2}}+\frac{2\alpha\eta\sigma
P}{(1+P)^{2}}+2\alpha\eta P-\Omega^{2}/2\right).
\end{equation} If the wave number of the perturbation becomes complex, then the instability will set in with exponential growth of the
perturbation field $a(\xi,s)$ resulting in the filamentation of the
broad beam into a number of filaments. Thus, the propagating broad
optical beam will be unstable. The growth rate $g(\Omega) (=2Im(K))$ of the MI is obtained as
\begin{equation}\label{e43}
g(\Omega)=\sqrt{2}\Omega\left(\frac{2g\beta\sigma
P}{(1+\sigma)(1+P)^{2}}+\frac{2\alpha\eta\sigma
P}{(1+P)^{2}}+2\alpha\eta P-\Omega^{2}/2\right)^{1/2}.
\end{equation} \subsection{Gain Spectrum of Instability }
MI gain  is achievable in a photovoltaic
PR (PVPR) crystal for a given range of frequency and  specified range of values of
$g,\sigma,\eta,\alpha,\beta$ and  $P$.  Parameters  $\eta$ and  $\sigma$  are always positive, while $\alpha$ and $\beta$ can be both positive or
negative depending on the media and the polarity of the external bias field. In an unbiased PVPR media
$( E_{a}=0\ i.e.,\beta=0)$ , thus K is always positive  if $\alpha< 0$. Hence, usually in such media, modulation instability
cannot set in. However, in such media  with the application of external field of appropriate magnitude and
polarity MI can set in and grow as long as

\begin{equation}\label{e44}
\frac{2g\beta\sigma P}{(1+\sigma)(1+P)^{2}}> 2|\alpha|\eta
P\left(1+\frac{\sigma}{(1+P)^{2}}\right)+\frac{\Omega^{2}}{2}.
\end{equation} Therefore, with the application of external electric field, it is
possible to initiate modulation instability and control the growth
of the instability in those media where MI was hitherto
prohibited. On the other hand, if for a given value of beam power MI growth rate
is finite in an unbiased PVPR medium with positive value of $\alpha$
, then the instability growth rate can be enhanced or decreased with
the application of external electric field of appropriate magnitude
and polarity. Or in other words, growth rate of the instability can
be controlled with the application of external field of chosen
polarity and magnitude. MI can takes place only
below the critical frequency $\Omega_{c}$ which is given by
\begin{equation}\label{e45}
\Omega_{c}=\pm\left[\frac{g\beta\sigma
P}{(1+\sigma)(1+P)^{2}}+\frac{\alpha\eta\sigma
P}{(1+P)^{2}}+\alpha\eta P\right]^{1/2}.
\end{equation} The instability is most efficient and reaches its maximum at $g=g_{m}$ when $\Omega=\Omega_{m}$, where $g_{m}= \frac{\Omega_{c}^{2}}{2}$  and $\Omega_{m}=\frac{\Omega_{c}^{2}}{\sqrt{2}}$. To examine the MI growth, we take a typical Cu:KNSBN crystal. Intermediate energy level is included in
Cu:KNSBN crystal and photovoltaic field is in the direction of optic axis. At $\lambda_{0}=0.5\mu m$, crystal parameters are $n_{e}=2.27,\ E_{p}=2.8\times10^{6} Vm^{-1}$, $\eta=1.5\times10^{-4},\sigma=10^{4}$,  $r_{33}=200\times10^{-12}m/V$. The scaling parameter  $x_{0}=10\mu m$,  $\alpha=117.3$  and $g=1$. We  take three different values of $E_{a}$, in particular, $E_{a}=-2\times 10^{6}
Vm^{-1}$, $0$ and $2\times 10^{6} Vm^{-1}$ which corresponds to
$\beta=-83.79,0$ and $ 83.79$ respectively. Figure (1) depicts
the MI gain spectrum $g(\Omega)$ as a function of perturbation
frequency $\Omega$. The variation of $\Omega_{m}$ with $P$ for three values of $\beta$ has been
depicted in figure (2). Initially $\Omega_{m}$ increases with the increase in the
value of $P$ then decreases with the increase in $P$.  In order to examine the role
played by  $\alpha$ on the growth of the instability,
we have demonstrated in figure (3) the variation of maximum growth
$g_{m}$ with the normalized beam power for three different value of
$\alpha$.  As expected, a higher value of $\alpha$ enhances the
growth of the instability. In conclusion, with the application of
external electric field, it is possible to initiate modulation
instability and control the growth of the instability in those media
where MI was hitherto prohibited. \section{Conclusion } A brief review of some selected developments in the field of optical spatial solitons in
PR media has been presented. Underlying mechanism responsible for the formation of solitons have been discussed for both single and two-photon PR media.
Vector solitons, particularly, incoherently coupled solitons due to single photon and two-photon
PR  phenomena have been highlighted. Existence of some missing solitons pointed out. Modulation instability which is a precursor to soliton formation has been also considered. Important applications of PR solitons have been highlighted.


\section*{Acknowledgment}

This work is supported by SAP programme of University Grants
Commsion (UGC), Government of India. One of the authors SK would like
to thank UGC for this.\\\

\newpage

\begin{figure}
\centering \scalebox{0.5}
{\includegraphics[height=10in,width=10in]{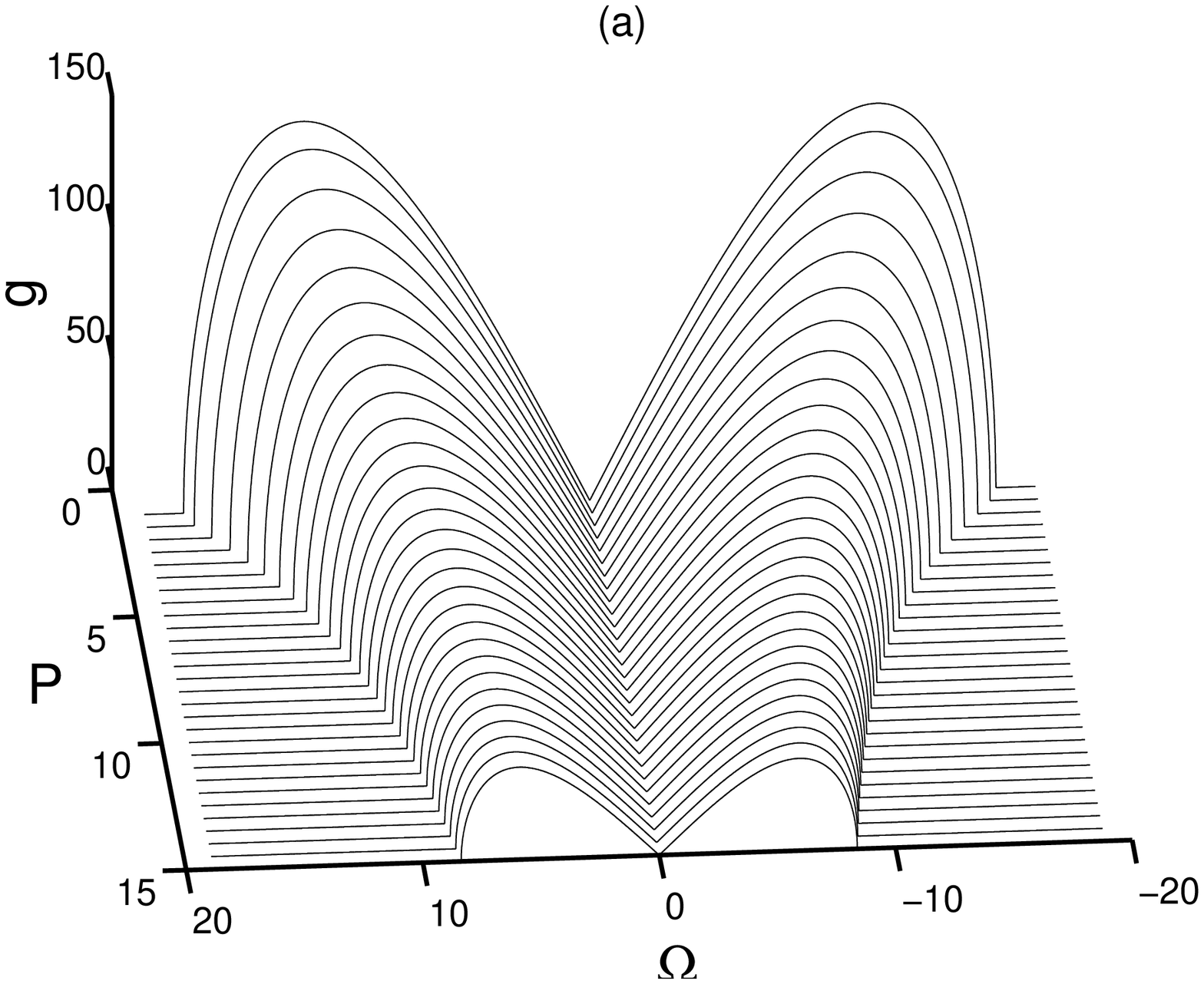}}
\caption{MI gain spectrum $g(\Omega)$ as a function of perturbation frequency $\Omega$} \label{fig:fig1}
\end{figure}

\newpage

\begin{figure}
\centering \scalebox{0.5}
{\includegraphics[height=10in,width=10in]{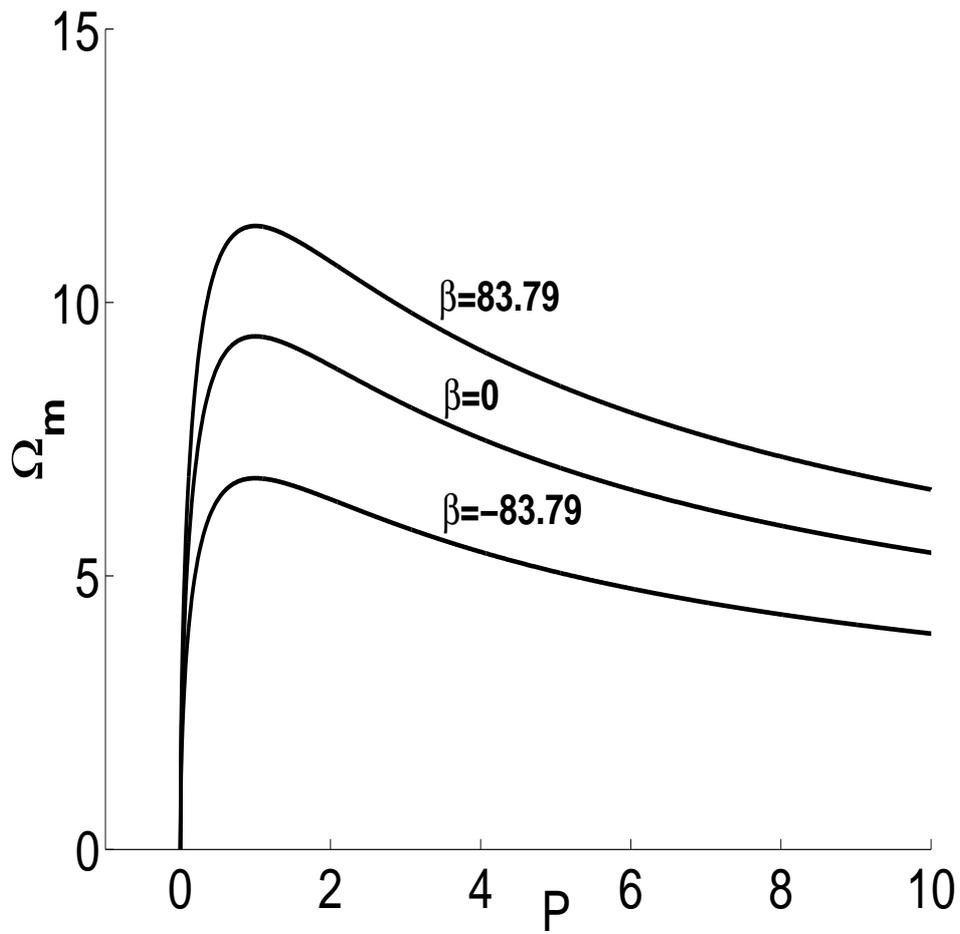}}
\caption{Variation of $\Omega_{m}$ with $P$ for three values of $\beta$.} \label{fig:fig2}
\end{figure}

\newpage
\begin{figure}
\centering \scalebox{0.5}
{\includegraphics[height=10in,width=10in]{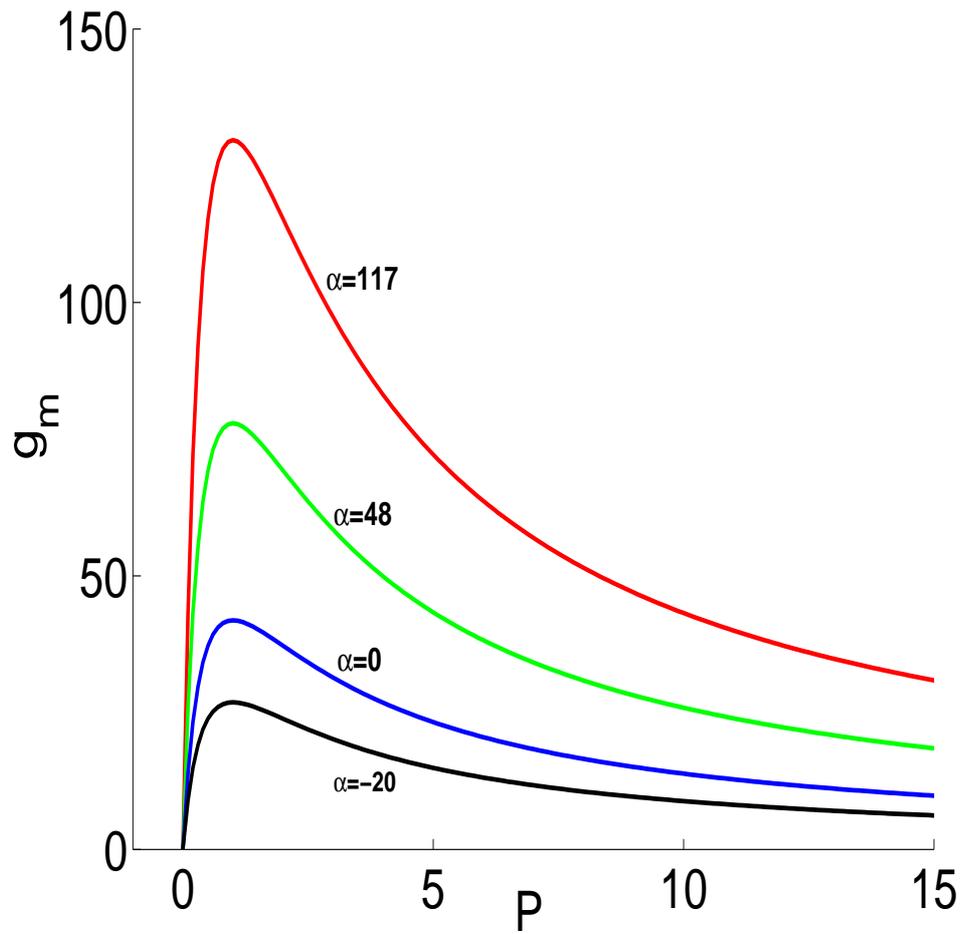}}
\caption{Variation of maximum growth
$g_{m}$ with the normalized beam power for  different values of $\alpha$.} \label{fig:fig3}
\end{figure}

\end{document}